\documentclass[preprint,merge]{aastex631}
\usepackage{amsmath}
\newcommand\Nf{N_\mathrm{f}}
\newcommand\Nt{N_\mathrm{t}}
\newcommand\Son{S^\mathrm{on}}
\newcommand\Soff{S^\mathrm{off}}
\newcommand\rhodif{\rho_\mathrm{dif}}
\newcommand\rhoref{\rho_\mathrm{ref}}
\newcommand\rLC{r_\mathrm{LC}}
\newcommand\tdif{t_\mathrm{dif}}       
\newcommand\tref{t_\mathrm{ref}}       
\newcommand\win[1]{$\text{w}_{#1}$}  
\newcommand{\nb}[1]{{#1}}
\newcommand\bms[1]{{#1}} 
\shorttitle{Interstellar interferometer}
\shortauthors{Popov et al.}
\begin{document}
\title{Technical constraints on interstellar
  interferometry and spatially resolving the pulsar magnetosphere} 
\correspondingauthor{M.~S.\ Burgin}
\email{mburgin@asc.rssi.ru}
\author[0000-0001-7931-646X]{M.~V.\ Popov}
\affiliation{Lebedev Physical Institute, Astro Space Center, Profsoyuznaya 84/32, Moscow, 117997, Russia}
\author{N.\ Bartel}
\affiliation{York University, 4700 Keele St., Toronto, ON M3J 1P3, Canada}
\author{A.~S.\ Andrianov}
\affiliation{Lebedev Physical Institute, Astro Space Center, Profsoyuznaya 84/32, Moscow, 117997, Russia}
\author[0000-0002-0579-2938]{M.~S.\ Burgin}
\affiliation{Lebedev Physical Institute, Astro Space Center, Profsoyuznaya 84/32, Moscow, 117997, Russia}
\author[0000-0002-3405-2795]{E.~N.\ Fadeev}
\affiliation{Lebedev Physical Institute, Astro Space Center, Profsoyuznaya 84/32, Moscow, 117997, Russia}
\author{A.~G.\ Rudnitskiy}
\affiliation{Lebedev Physical Institute, Astro Space Center, Profsoyuznaya 84/32, Moscow, 117997, Russia}
\author{T.~V.\ Smirnova}
\affiliation{Lebedev Physical Institute, Pushchino Radio Astronomy Observatory, Pushchino 142290, Moscow region, Russia}
\author{V.~A.\ Soglasnov}
\affiliation{Lebedev Physical Institute, Astro Space Center, Profsoyuznaya 84/32, Moscow, 117997, Russia}
\author{V.~A.\ Zuga}
\affiliation{Lebedev Physical Institute, Astro Space Center, Profsoyuznaya 84/32, Moscow, 117997, Russia}
\begin{abstract}
\nb{Scintillation} of pulsar radio signals caused by the interstellar medium can in
principle be used for interstellar interferometry. Changes of the dynamic
spectra as a function of pulsar longitude were in the past interpreted as having
spatially resolved the pulsar magnetosphere.  Guided by this prospect we used
VLBI observations of PSR B1237+25 with the Arecibo and Green Bank radio
telescopes at 324 MHz and analyzed such \nb{scintillation} at separate longitudes of
the pulse profile.  We found that the fringe phase characteristics of the
visibility function changed quasi-sinusoidally as a function of longitude.
Also, the dynamic spectra from each of the telescopes shifted in frequency as a
function of longitude. Similar effects were found for PSR B1133+16.  However, we
show that these effects are not signatures of having resolved the pulsar
magnetosphere. Instead the changes can be related to the effect of low-level
digitizing of the pulsar signal. After correcting for these effects the
frequency shifts largely disappeared. Residual effects may be partly due to feed
polarization impurities. \nb{Upper limits for the pulse emission altitudes of PSR B1237+25 would likely be well below the pulsar light cylinder radius.} In view of our analysis we think that observations
with the \nb{intent} of spatially resolving the pulsar magnetosphere need to be
critically evaluated in terms of these constraints on interstellar
interferometry.
\end{abstract}
\keywords{scattering --- pulsars: individual B1237+25 -- techniques}
\section{Introduction} \label{sec:intro}
Scattering of radio waves by inhomogeneities of the
interstellar plasma causes angular broadening of the source
image, distortion of radio spectra, and \nb{scintillation}.
Refractive \nb{scintillation} are slow and occur on time scales
of $\tref$ and diffractive \nb{scintillation} are fast and occur
on time scales of $\tdif$, with corresponding spatial scales
of $\rhoref$ and $\rhodif$.  Fluctuations in the density of
free electrons in the interstellar plasma produce variations
in the index of refraction. Pulsars observed through such a
medium produce a diffractive pattern in the plane of the
observer. \nb{The medium can be seen as a lens or an interferometer with an extremely large baseline and therefore, depending on the distance of the medium from the pulsar, providing extremely high spatial resolution.}
\nb{Using interstellar interferometry it is therefore possible in principle to probe the sizes of 
pulse emission regions, as well as their separations and relative motions with a spatial resolution in most cases much better than a light cylinder radius, $\rLC=cP/2\pi$, with $c$ the speed of light and $P$ the pulsar period.} 
Early discussions and studies were reported by
\citet{1968Natur.218..920S} and \citet{1970PhDT.......113L}. 
 The theory of interstellar plasma lensing was further developed by, e.g.,
 \citet{1977ARA&A..15..479R,CordesPidweb1986,1998ApJ...505..928G}.

First attempts to resolve the
pulsar magnetosphere  were made by
\citet{Backer1975} and \citet{cwb1983} by comparing scintillation patterns of
different pulse profile components. No evidence of
independent scintillation was found resulting in upper
limits of the corresponding emission regions of as small as
$1.2\rLC$ \nb{or $3\times10^4$  km for PSR B0823+26 and $1\times10^4$ km 
or, with particular assumptions, even $1\times10^3$ km for PSR B0525+21, respectively.} 

First phase sensitive attempts of resolving the pulsar magnetosphere were done
by \citet{Bartel+1985} with VLBI and gating. These authors discussed the effects
of polarization impurities of the feeds on measurements of \nb{visibility} phase as a function
of \nb{pulse} longitude \nb{of PSR B0329+54} and therefore on the separation of emission regions. They derived
stringent limits for the feeds so that the bias due to the changing polarisation
characteristics across the pulsar's profile would be reduced for sensitive
fringe phase measurements along pulsar longitude. Again, no evidence for having
resolved the pulsar magnetosphere was found.

The first apparently clear evidence for having resolved with interstellar
interferometry the magnetosphere of a pulsar was reported by \citet{wol_cor1987}
by having made use of occasionally occurring refractive scintillation. A strong
refraction event can split the image into two or multiple
subimages. \citet{wol_cor1987} observed periodic structure in the dynamic
spectrum of PSR B1237+25 with the Arecibo telescope (AR) at 430 MHz. They
interpreted that structure as an interference pattern formed when two radiation
beams having different paths through the interstellar medium (ISM) intersect in
the observer plane.  \nb{They detected an  asymmetric non-monotonic
smooth fringe phase
variation of the periodic structure in the dynamic spectra as a function of
pulse longitude and estimated a typical transverse separation between the
emission regions of $\sim10^3$ km for a screen at half distance between the pulsar and Earth.  For a magnetic dipole field such separation would place the emission regions at $\rLC$. Only a screen much closer to the pulsar would lower the emission altitude substantially. The authors also discussed an alternative of a very distorted dipole field for a reduced altitude of the emission region.}

Very similar results were reported for PSR B1133+16 by \citet{GuptaBR1999} using
the Ooty radio telescope at 327 MHz. They also used multiple imaging during a
refractive event and again found a non-monotonic 
fringe phase variations across
the pulse profile.  They inferred a \nb{minimum} separation of the emission regions of the
leading and trailing parts of a pulse of 
$3\times10^7$~m corresponding to a minimum
emission height of $\geq 2.6\times10^3$~km.

Further apparently successful attempts to spatially resolve the emission region
were reported by several authors. 
\cite{SS1989} analyzed dynamic diffractive scintillation spectra as a function
of longitude and found through cross-correlation analysis lags in time and
frequency indicating separations of the corresponding emission regions of
$3\times10^2\,$km, which for a dipole magnetic field corresponds to \nb{an emission height of} $0.08\rLC$. In
a similar analysis using however the cross-correlation peak degradation as a
function of longitude separation, \citet{smirnova1992} and \citet{SSM1996} found
for some other pulsars, including PSR B1237+25, in contrast, emission heights
close to $\rLC$. \nb{Similarly, focusing on giant pulses of the Crab pulsar, \citet{Main2021} found with diffractive scintillation for the main pulse and interpulse sizes and separations of approximately $\rLC$, although it should be noted that in this case $\rLC$ is with ~1,600 km relatively small.}

\nb{On the other hand, \citet{2012ApJ...758....8J} used diffractive scintillation of
the Vela pulsar and measured the size of the emitter to be  $<4$~km and its
height to be $<3.4\times10^2$~km, \nb{much smaller than $\rLC$.}} 
Also \citet{Pen} following \citet{Brisken+2010} \nb{reduced VLBI observations of PSR
B0834+06 with primarily AR and the Green Bank Telescope (GB)  with 4-level digitization at 327 MHz and then used} VLBI imaging of its scattering speckle pattern to
measure the changing phase response on the scattering screen as a function of
pulse longitude. \nb{The authors found a phase change of $\sim0.01$~rad over an $\sim18$~ms longitude range of the pulse profile. They
interpreted it as a deflection or an apparent motion of only 18 km or 
$\sim1000\,\text{km\,s}^{-1}$ of a small emission region at an altitude of a few hundred km,} effectively performing high-precision astrometry
with \nb{an angular} resolution of 50 picoarcseconds.

\nb{Summarizing, although there have been efforts for almost half a century to resolve the pulsar magnetosphere with sometimes apparently extremely high spatial resolution, no consistent picture as to the magnitude of the displacements of the emission regions along the pulse profile or the emissions' altitude, near the pulsar or close to $\rLC$, has yet emerged.

Inspired by these results from interstellar interferometry we focused on the multicomponent pulsar B1237+25 and used VLBI observations with the AR and GB telescopes as well as the respective observations with the single antennas.
Our goal was to probe the scintillation properties and determine any changes as a function of pulsar longitude to derive the corresponding emission regions' separations or their upper limits. In Section \ref{sec:Observations} we describe our observations and the primary reduction of our data. In Section \ref{sec:phase_freq_shift} we show the visibility phases from the VLBI observations as they change as a function of pulse longitude and compare the results with the corresponding frequency shifts of the dynamic spectra for each of the single telescopes. In Sections \ref{sec:causes} and \ref{sec:reduction} we explain why our changes in the scintillation properties are not indicative of having resolved the pulsar magnetosphere but in contrast can be largely if not completely attributed to effects of low-level digitization of the data. In Section \ref{sec:discussion} we critically discuss our findings in view of previous reports of having resolved the magnetosphere, emphasize the technical constraints that need to be considered for interstellar interferometry and draw conclusions about the physics of the emission regions and the distance of the scattering screen for PSR B1237+25. In Section \ref{sec:conclusion} we give a summary of our results and our conclusions.}

\section{Observations and primary data reduction}   \label{sec:Observations}
\nb{We made VLBI
observations} of PSR B1237+25 with AR and GB at
324 MHz with a bandwidth, $B =16$~MHz at right (RCP) and left-circular (LCP)
polarization.  The observations were made in the context of the Radioastron
space-VLBI science program AO-5.  We used two observing sessions separated by
two months for the PSR B1237+25, and one observing session for the PSR B1133+16, each 2~h long.  General information about the sessions is given in
Table \ref{tab:obs_table}.  All sessions consisted of six 1160~s long recording
scans separated by 30~s gaps.  During each scan, 840 pulsar periods with
$P = 1.38$~s were used for our analysis of data for PSR B1237+25. The data were processed at Astro Space
Center with the ASC software correlator \citep{2017JAI.....650004L} with gating
and incoherent dedispersion applied.  We used 512 channels at the correlator,
providing a frequency resolution of 31.25~kHz.
\begin{deluxetable*}{lcchlRCR} 
\tablecaption{List of observations. \label{tab:obs_table}}
\tablehead{
\colhead{PSR} & \colhead{Date}  & \colhead{Time} &
\nocolhead{
Baseline
} & \colhead{Stations} & \colhead{$\Delta f_{1/e}$} & \colhead{$\tdif$} & \colhead{$\Delta \tau_{1/2}$}
\\
 & (yyyy mm dd) & (hh:mm -hh:mm) & 
\nocolhead{
projection
} & &
\colhead{(MHz)} & \colhead{(s)} & \colhead{(ns)}
\\
\colhead{(1)} & \colhead{(2)} & \colhead{(3)} &
&\colhead{(4)} &\colhead{(5)} &\colhead{(6)}
&\colhead{(7)} 
\\
}
\startdata
\nb{B1237+25}             & 2017 12 22 & 10:00 - 12:00 & 20.7\,D_\Earth  & AR, GB, WB   & 1.70-4.40   & 345\pm20 &65\pm~5\\
\nb{B1237+25}             & 2018 02 26 & 05:30 - 07:30 & 13.7\,D_\Earth  & AR, GB       & 0.71-1.22     & 250\pm10 & 260\pm10\\ 
\nb{B1133+16}             & 2018 12 17 & 09:00 - 11:00 & 22.0\,D_\Earth  & AR, GB, WB   & 1.50-2.30           & 80.4\pm4& 320\pm20\\
\enddata
\tablecomments{
Columns are as follows:
(1) \nb{Pulsar observed.}
(2) Date of observations \nb{related to observing codes rags29c, rags29j, and raks24e for the first, second and third epoch, respectively.}
%
(3) Time span of observations for six scans. 
Each scan is 1160 s long which
corresponds for PSR B1237+25 to 840 pulsar periods and for PSR B1133+16 to 976 pulsar periods.
(4) Radio antennas scheduled for the observations, AR -- Arecibo, GB -- Robert
C. Byrd Green Bank Telescope, WB -- Westerbork. We report here only results for
AR and GB in VLBI mode and single antenna mode. 
(5) Decorrelation bandwidth as half-width at 1/e of the maximum of the frequency
section of the 2-dim autocorrelation function (ACF) at zero time lag measured
with AR for window, \win{4}, of the pulse profile (see Figure ~\ref{fig1}). The
ranges refer to the minimum and maximum values for the six scans for each of the
two epochs. 
(6) Diffractive scintillation time.
(7) Half-width at half maximum of the visibility magnitude obtained in
sub-window \win1 for one scan.  Uncertainties in (6) and (7) are statistical
standard errors. 
}
\end{deluxetable*}

\begin{figure*}[htb]
\includegraphics{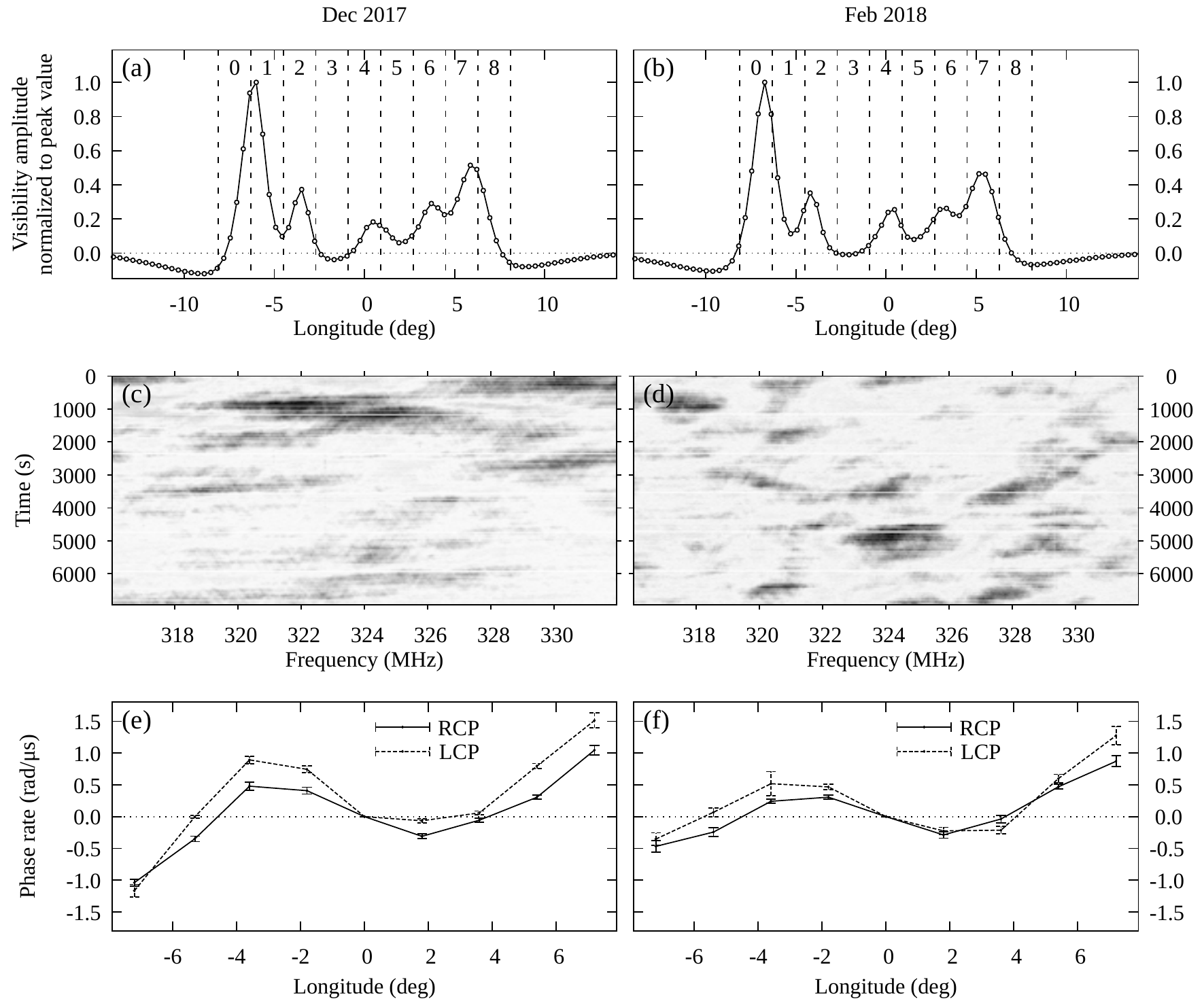}
\caption{ 
Pulse profiles of PSR B1237+25 observed at RCP (panels a, b), dynamic spectra of
diffractive scintillation at RCP (panels c, d), and rate of change of the
visibility phases as a function of pulsar longitude for both senses of
polarization (panels e, f), each averaged over six scans. Left and right panel
columns are for our two observing epochs, 2017 December 22 and 2018 February 26,
respectively.
Pulse profiles in (a, b) are computed as the visibility magnitude obtained
by the correlator at zero baseline for AR with the off-pulse levels
subtracted.  The vertical dashed lines 
indicate the nine on-pulse sub-windows, \win0 to \win8, used as
gates for the correlation and for our analysis.
Dynamic spectra in (c, d) are the averages for AR over
the full on-pulse windows and shown on normalized linear
gray-scale. The black and white regions represent the
maxima and minima of the power density, respectively.
Phase rates, in (e, f), are the derivatives of the AR-GB VLBI visibility phases \nb{from strong pulses}
with respect to delay, $\frac{d(\varphi_k - \varphi_4)}{d\tau}$ for the nine pulse windows with
$0\leq k\leq8$. Uncertainties are standard errors. They were derived from the $1\sigma$
uncertainty from the least-squares fit of the phase rates in the approximately
$\pm$ 125 ns central delay region for each of the six scans and then divided by
$\sqrt6$.
\label{fig1}
}
\end{figure*}

First, we computed the average pulse profile and dynamic spectrum for AR for
each of the two observing sessions. The pulse profiles are shown in Figures
~\ref{fig1}a and \ref{fig1}b.  The width of the pulse
window is $\sim65$~ms, corresponding to a longitude range of $\sim16$~deg.  Note, the
slight false decrease of intensity at the leading and trailing part of the
profile \nb{which we will discuss in Section \ref{sec:reduction}}.  The dynamic spectra are shown in Figures~\ref{fig1}c and
\ref{fig1}d.  The spectra are similar to what would be expected
under regular, diffractive, scattering conditions. No quasi-periodic modulation
as a consequence of multiple imaging due to refraction is observed.

Second, we divided the pulse window into nine sub-windows, w$_k$, with $0\leq k\leq8$,
each $\sim7$~ms wide (see Figures \ref{fig1}a and
\ref{fig1}b).  We also selected three off-pulse sub-windows, w$_k$,
with $9\leq k\leq11$, of the same duration.  Then we computed with the correlator for
each window the complex cross-spectrum for the baseline AR-GB, $S^{AR-GB}_k(f_i,
t_j)$, and the auto-spectrum, $S_k (f_i, t_j)$ (dynamic spectrum), for each of
the two radio telescopes.  Each dynamic spectrum consists of $ \Nf\Nt $ values,
with $1 \leq i \leq \Nf$ and $1 \leq j \leq \Nt$ where $ \Nf $ is the number of frequency
channels covering the frequency range in the bandpass from 316 to 332 MHz, and $
\Nt $ the number of spectra in a given set of observations.  We corrected the
dynamic spectrum $S_k (f_i, t_j)$ in each sub-window, w$_k$, for the background
baseline as follows:
\begin{equation}        \label{eq:cal_dyn}
 S_k (f_i, t_j) = \Son_k (f_i, t_j) -\Soff(f_i, t_j). 
\end{equation}
Here, $ \Son_k $ is the spectrum obtained in each of the sub-windows, \win0 to
\win8, and $ \Soff $ is the spectrum average over the off-pulse windows, \win9
to \win{11}.  The individual spectra were averaged over four pulse periods to
smooth out the intensity fluctuations from pulse to pulse, reducing $\Nt$ to 210
but keeping $\Nf= 512$.  

Third, we computed the two-dimensional
cross-correlation functions between the dynamic spectra of each of the
sub-windows and the dynamic spectrum of the sub-window in the center of the
pulse, w$_4$, which served as a reference.
\begin{equation}      \label{eq:defACF}
CCF_{4k} (\Delta f,\Delta t)= {\frac{1}{(\Nf-i) (\Nt-j)}} 
      \sum_{i=0}^{\Nf-i}\sum_{j=0}^{\Nt-j}\Delta S_{4}(f_i,t_j)\Delta S_{k}(f_i+\Delta f,t_j+\Delta t).
\end{equation}
Here $\Delta f={\frac{B}{\Nf}}i$ and $\Delta t=4Pj$ are frequency and time lags, and $\Delta
S_{k}(f,t)=S_{k}(f,t)-\langle S_{k}(f,t) \rangle$ with
\begin{equation}                          \label{eq:Savrg}
\langle S_{k}(f,t) \rangle={\frac{1}{\Nf \Nt}}
  \sum_{i=1}^{\Nf}\sum_{j=1}^{\Nt}S_{k}(f_i,t_j)\,.
\end{equation}
After the primary data reduction we proceeded to the main analysis.
\section{Phase and frequency shifts as a function of pulsar longitude}
\label{sec:phase_freq_shift}
\subsection{Phase shifts of the VLBI visibility functions } \label{sec:phase}
\begin{figure}[htb]
\includegraphics{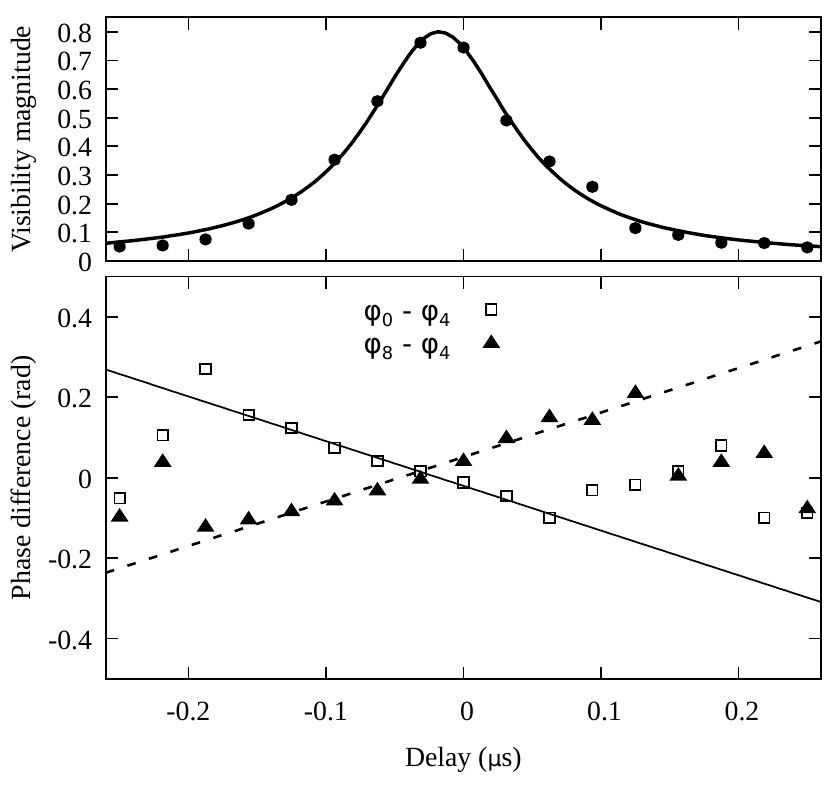}
\caption{
 Upper panel: Average AR-GB visibility magnitude as a function of interferometer
 delay for PSR B1237+25 obtained in sub-window \win1 for one scan (10:00) on
 2017 December 22. The solid line shows a Lorenzian, fit to the data. It has a
 half-width-at-half-maximum (HWHM) of 65~ns.  Lower panel: Corresponding average
 differences of the visibility phases between the leading (\win0) and trailing
 (\win8) sub-windows relative to the phases for sub-window, \win4. The phases
 were corrected for $2\pi$-ambiguities.  Only RCP pulses with the highest SNR were
 used. The tilted lines are least-squares fits to the aligned phases in the
 central region of the visibility magnitude curve. Uncertainties are smaller or
 approximately equal to the symbol sizes. Phases outside the central region have
 much larger errors.
\label{fig2}
}
\end{figure}

For every single pulse and every sub-window, with $0 \le k \le 11$, we computed the
complex visibility function $V_{k}(\tau)$ as the inverse Fourier transform of
$S^{AR-GB}_k(f)$ with the sampling interval in interferometer delay, $\tau$, being equal to
31.25~ns.  We used only visibilities for the AR-GB baselines.  As an example we
give the average of the visibility function magnitude for one scan and for
window \win1 in Figure \ref{fig2}.
Then we investigated the phases, $\phi_k$ of $V_k(\tau)$ relative to $\phi_4$ of $V_4(\tau)$
as a function of longitude.  We selected only strong pulses with the
signal-to-noise ratio (SNR) in both the selected sub-window, \win{k}, and
sub-window \win4 being $>$5 with respect \nb{to the} off-pulse window, \win{11}.  For these
pulses we computed for each sub-window in the pulse profile the phase relative
to the phase in sub-window, \win4, ($\varphi_k - \varphi_4)$.  We analyzed data from six
scans and obtained for every scan and every window in general several hundreds
of phase differences from our strong pulses.  

As an example, we present the
average phase differences for the leading ($\varphi_0 - \varphi_4)$ and the trailing ($\varphi_8 -
\varphi_4)$ sub-windows as a function of 
$\tau$ in
Figure~\ref{fig2}. In the approximately $\pm 125\, \text{ns}$ delay
window the curves of phase differences can be approximated by straight lines
with a fitted slope of $\frac{d(\varphi_k - \varphi_4)}{d\tau}$ for k=0 and 8.  As can be
seen in Figure~\ref{fig2}, the slopes are very different for the
leading and trailing windows relative to window 4. In general, these slopes vary
across the pulse window. In Figures~\ref{fig1}e and
\ref{fig1}f we plot the values of the varying slopes and their
standard errors from the fit for each sub-window with respect to the value for
window, \win4, for the two polarizations and the two days of observations.  

The variations are smooth, highly significant and appear to be
quasi-sinusoidal \nb{with an upward trend along longitude}. The patterns in Figures~\ref{fig1}e and
\ref{fig1}f are very similar for the two polarizations and for the
two days.  We note that our pattern of the VLBI phase rate versus pulse
longitude is also very similar to the comparable pattern of phase versus
longitude presented for PSR B1237+25 by \citet{wol_cor1987} and for PSR B1133+16
by \citet{GuptaBR1999}. However, we will show that in our case this pattern is
not an indication of having resolved the pulsar magnetosphere.  The variation of
the derivative of phase along longitude can be converted to a frequency shift through the relation $|\Delta f|=\frac{1}{2\pi}|\frac{d\varphi}{d\tau}|$.  From, e.g.,
Figure~\ref{fig1}e we obtain the total change of the LCP phase
derivative across the profile from \win0 to \win8 of $\sim2.65\,\text{rad}/\mu s$
which, with the shifting property of the Fourier transform, corresponds to a
shift in frequency of the dynamic spectra to lower frequencies by an amount of
$\sim$-420 kHz.
\subsection{Frequency shifts of the dynamic spectra}
\label{sec:DSP} 
\begin{figure*}[htb!]
\includegraphics{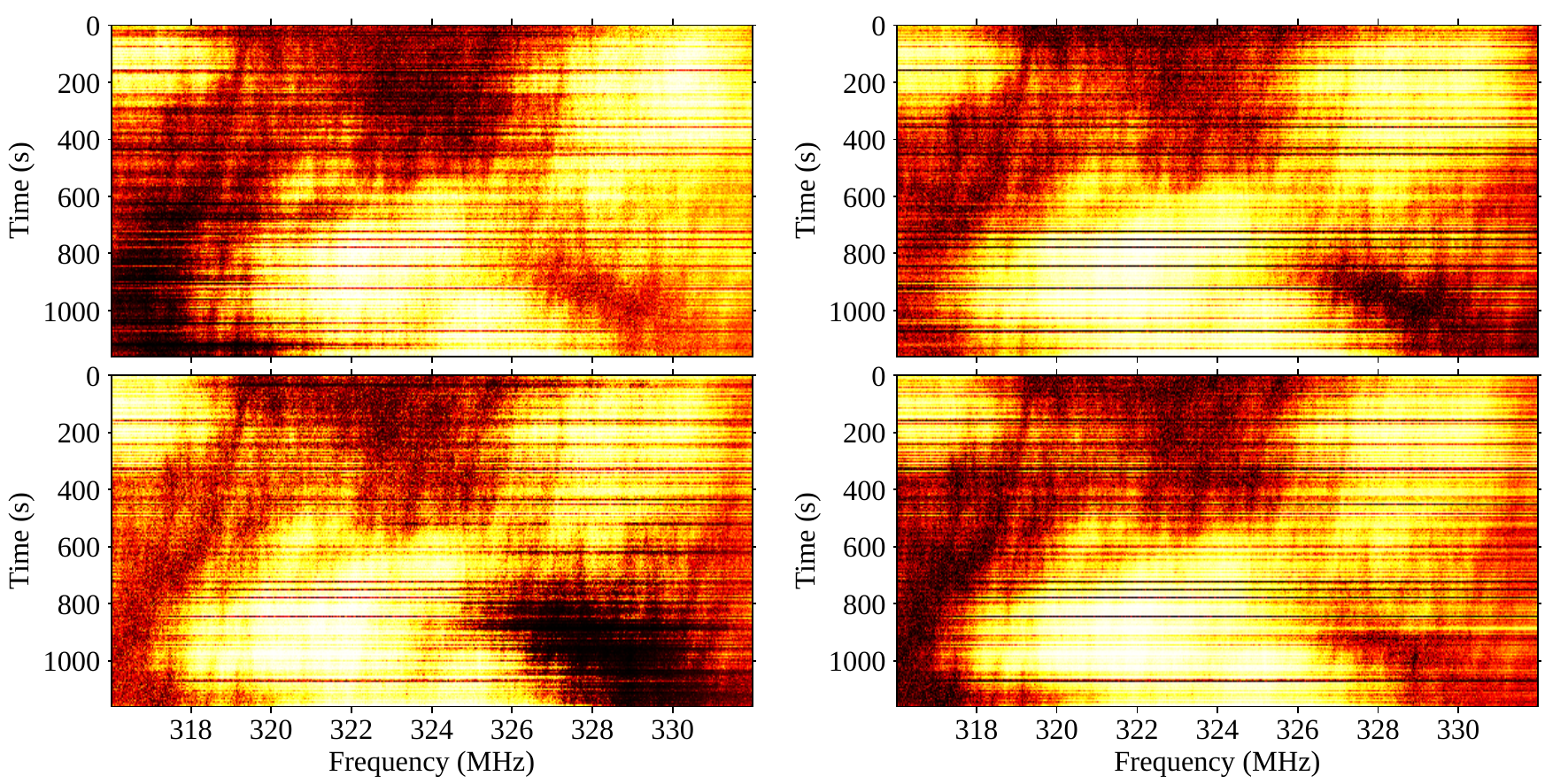}
\caption{The dynamic spectra in longitude windows w$_0$ (top) and w$_8$ (bottom)
  observed at AR in LCP on 2017 December 22 (scan 1).  The left panels
  show the dynamic spectra before correction. A shift of about 1 MHz toward
  lower frequencies from w$_0$ to w$_8$ can be seen.  The right panels show the
  corresponding spectra after correction as described in Section
  \ref{sec:causes}. The spectrum in w$_0$ is copied from the left side for
  better comparison. The spectrum in w$_8$ shows that the frequency shift has
  largely disappeared. \nb{For the difference of the time-averaged spectra and a discussion, see Subsection \ref{subsec:delay}.}  \label{fig3}
}
\end{figure*}

  We can also see the frequency shift in our dynamic spectra as a function of
  longitude for each of the two telescopes separately. In
  Figure~\ref{fig3} we show on the left side the dynamic spectra in windows
  \win{0} (top panel) and \win{8} (bottom panel). The latter one is slightly
  but clearly shifted toward lower frequencies with respect to the former. 

A more detailed presentation of the frequency shift is obtained with $CCF_{4k}(\Delta
f, \Delta t$), of the dynamic spectra magnitudes $S(f,t)$.  First, we determined the
decorrelation bandwidth, $\Delta f_{1/e}$ as the lag at 1/e of the maximum from the
frequency cross-section of the autocorrelation function, $CCF_{44}$, of the
dynamic spectra in \win4. We list the range of values for the six scans for each
of the two observing epochs in Table \ref{tab:obs_table} and list the individual
values in Table \ref{tab:shift_vs_decorr}.

\begin{figure*}[htb!]
\includegraphics{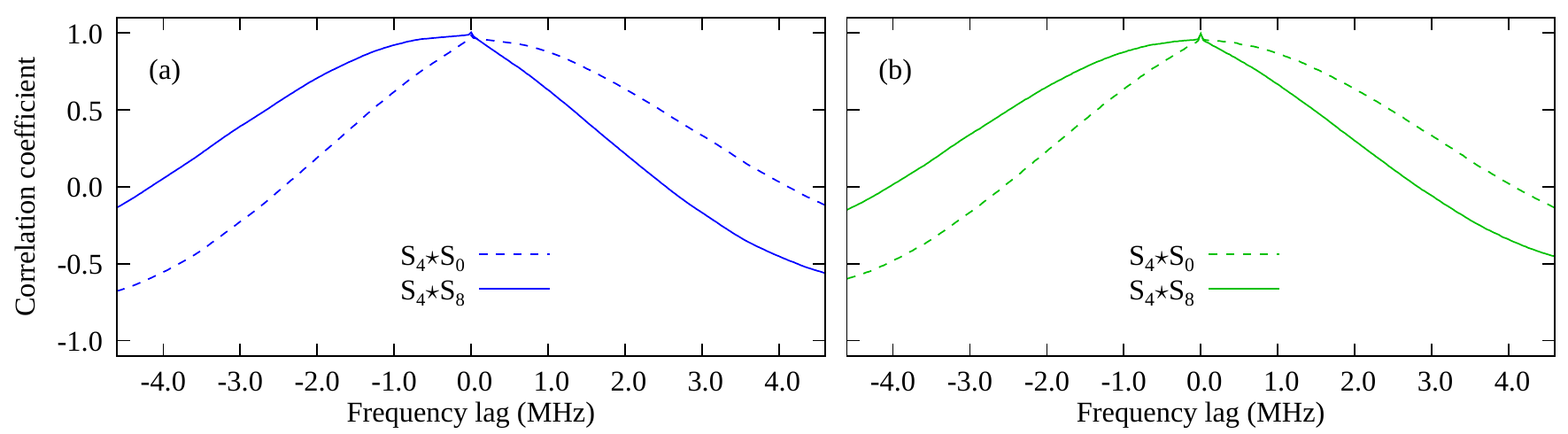}
\caption{Examples of frequency sections of the two-dimensional CCF($\Delta f, \Delta t$)
  functions for $\Delta t=0$ of dynamic spectra as a function of frequency lag,
  $\Delta f$. Panels (a) and (b) correspond to AR and GB, respectively.  The CCF's
  are computed for two pairs of dynamic spectra, one corresponding to the leading
  and the middle window (\win0-\win4, solid curves) and the other to the trailing
  and the middle window (\win8-\win4, dashed curves). The CCFs were normalized to
  have equal values at zero lag.  LCP data were used from scan 1 at epoch 2017
  December 22.
\label{fig4}}
\end{figure*}

Second, we determined the frequency shift of the
dynamic spectra in each of the sub-windows relative to the spectrum in \win4.
For this purpose we used the frequency sections of 
$CCF_{4k}(\Delta f, \Delta t$) for $\Delta t=0$.  As an example we take the dynamic spectra
from \win0 and \win8 from Figure~\ref{fig3} (left panels) and
cross-correlate them with respect to that of \win4. The corresponding functions
$CCF_{4k}$ with k=0 and 8 are plotted for AR and GB in
Figure~\ref{fig4}.
A shift of spectra in \win8 to lower frequencies with respect to spectra in
\win0 is clearly visible.

For an accurate determination of the frequency shift we accounted for the
asymmetry of the functions by fitting to them the function, $Y(x)$, with $x=\Delta f$
and $x_0$ as the frequency shift with
\begin{equation}     \label{eq:Yx}
Y(x)=A\exp{{\left(-\frac{|x-x_0|}{B}^{\alpha}\right)}}+C+D(x-x_0).
\end{equation}
Here, $C$, compensates for a possible baseline offset, D, accounts for the
asymmetry of the CCF and, $\alpha$, for the shape of the CCF.  We determined the
frequency shifts, $x_0$, for the dynamic spectra in each window relative to that
in \win4 for each of the six scans and each of the observing epochs.  The fitted
values of $\alpha$ are in the range of 1.5 to 1.8. The frequency shift values, $x_0$,
correspond to shifts of the spectra from the leading part of the pulse profile
in \win0 to the trailing part in \win8.

\begin{figure*}[htb!]
\includegraphics{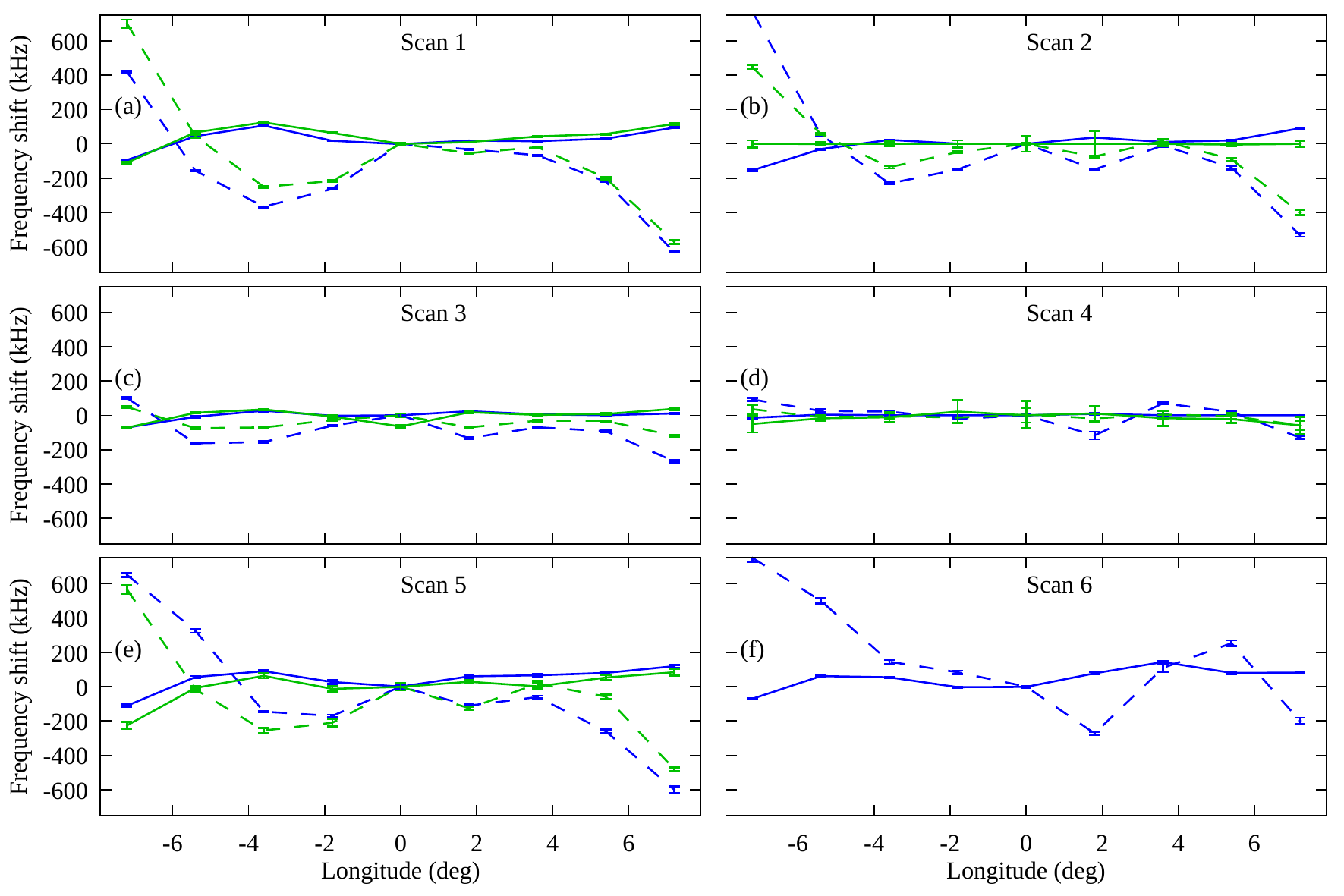}
\caption{Frequency shifts of dynamic spectra in \win0 to \win8 relative to
  spectrum in \win4.  Panels (a-g) present results for the six scans obtained in
  LCP at AR (\bms{blue} lines) and GB (green lines) on 2017 December 22. For scan 6 the
  GB data were not usable and are omitted. Each scan is 840 pulse periods
  long. Dashed lines correspond to original, uncorrected frequency shifts. Solid
  lines correspond to remaining relatively small frequency shifts after
  correction for distortion discussed in subsection
  ~\ref{subsec:delay}. Uncertainties are 1$\sigma$ statistical standard errors
  determined from the fit of eq.~\ref{eq:Yx}.
 \label{fig5}
}
\end{figure*}

\begin{figure*}[htb!]
\includegraphics{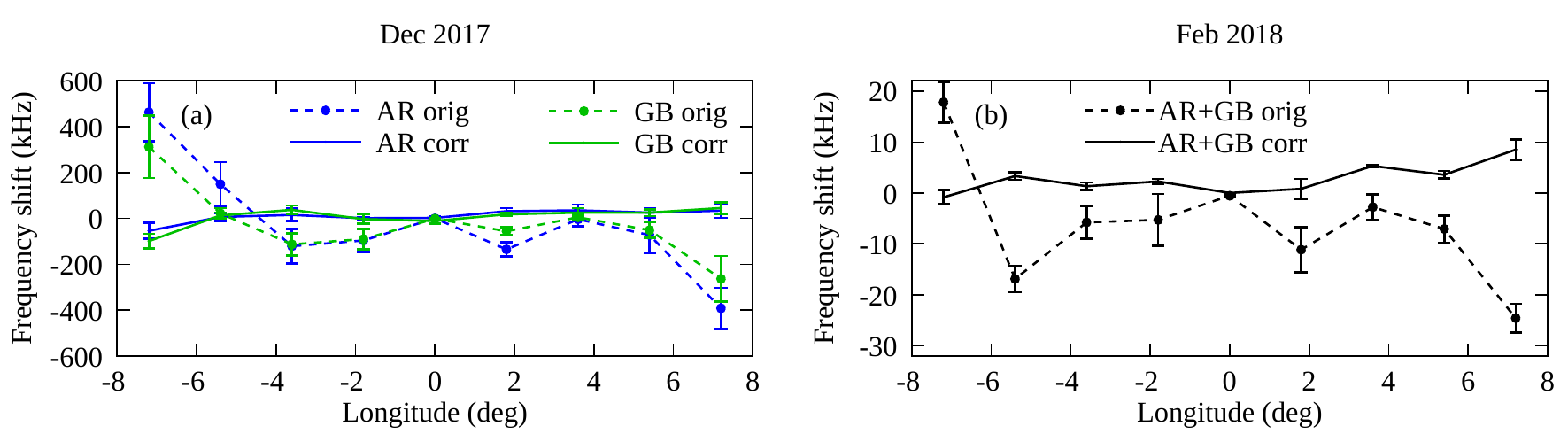}
\caption{Left panel: The averages 
  of the frequency
  shifts for 2017 December 22 for windows \win{0} to \win{8} relative to \win{4} for
  the six scans in Figure ~\ref{fig5}. Right panel: The
  corresponding frequency shifts for 2018 February 26 averaged over all scans,
  the two polarizations and the two telescopes.  Uncertainties are standard errors.
\label{fig6}}
\end{figure*}

We plot the frequency shifts with dashed lines for AR and GB for each of the
scans for the first epoch in Figure ~\ref{fig5} and plot the
averages from all six scans in Figure ~\ref{fig6}.  The frequency
shifts for our second epoch could be better inspected by averaging over all six
scans and the two polarizations and telescopes.  For comparison we plot these
averages in the right panel in the same Figure.

Figure ~\ref{fig6} shows that the quasi-sinusoidal modulation of
the frequency shift as a function of longitude is very similar for AR and GB on
2017 December 22. It is also similar to the shape of the modulation on 2018
February 26. However, the modulation amplitude is about 20 times smaller. We
found that this decrease in amplitude is related to a decrease in $\Delta f_{1/e}$.

\begin{figure}[htb!]
\includegraphics{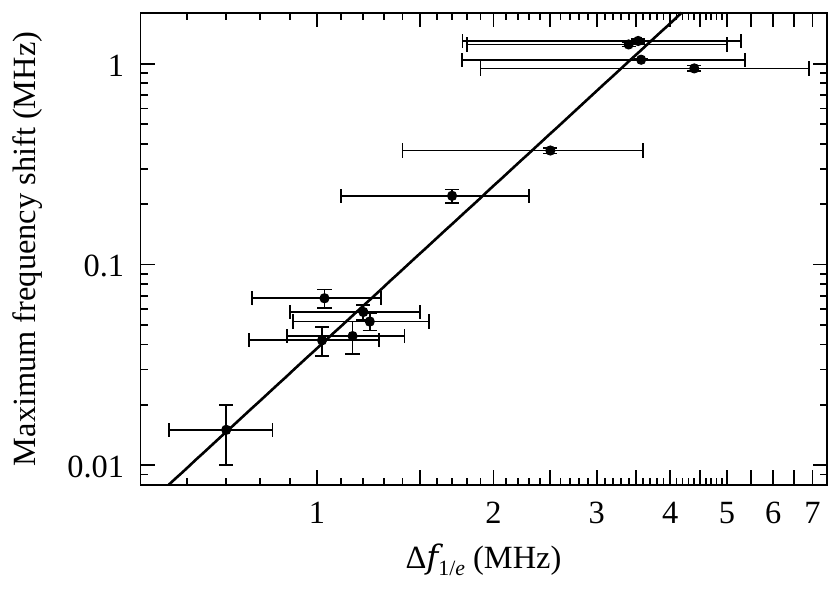}
\caption{Maximum frequency difference observed in AR LCP dynamic spectra in
  \win8 relative to \win0 as a function of decorrelation bandwidth, $\Delta f_{1/e}$,
  taken from Table~\ref{tab:shift_vs_decorr}. The shift is to lower frequencies
  for the dynamic spectra from the leading part of the pulse profile in \win8 to
  the trailing part in \win0. The data from Feb. 2018 February 26 are all in the
  lower left corner.
\label{fig7}
}
\end{figure}

In Table ~\ref{tab:shift_vs_decorr} we list the magnitudes of the maximum
frequency shifts as a function of $\Delta f_{1/e}$ for each of the six scans of the
two epochs and plot them in Figure ~\ref{fig7}. A weighted
least-squares fit of the maximum frequency shift, $M_{f-shift}$, to a
polynomial, $M_{f-shift}=A(\frac{\Delta f_{1/e}}{MHz})^b$, gives A=0.04$\pm$0.01 MHz
and b=2.7$\pm$0.2 with scaled uncertainties so that Chi-square per degree of
freedom, $\chi^2_{\nu} =1$.
\begin{deluxetable*}{lcRC} 
\tablecaption{Decorrelation bandwidth and freq. shift}
\tablehead{
\colhead{Date} & \colhead{Scan}  & \colhead{$\Delta f_{1/e}$} & \colhead{Max. freq. shift} 
\\
(yyyy mm dd)& (number) & (MHz) & (MHz) 
\\}
\startdata
2017 12 22  & 1&3.57\pm1.80 & 1.05\pm0.01\\
            & 2&  3.53\pm1.76 & 1.30\pm 0.03\\
            & 3&  2.50\pm1.10  & 0.37\pm0.01\\
            & 4&  1.70\pm0.60 &  0.22\pm0.02 \\
            & 5 & 3.50\pm1.60 & 1.25\pm0.03 \\
            & 6 & 4.40\pm 2.50 & 0.95\pm 0.03 \\
2018 02 26  & 1 & 1.15\pm 0.26 & 0.044 \pm 0.008 \\
            &  2 & 0.70 \pm 0.14 & 0.015\pm 0.005 \\
            &  3 & 1.02\pm 0.26  & 0.042\pm 0.007 \\
            &  4 & 1.03 \pm 0.26 & 0.068 \pm 0.007 \\
            &  5 & 1.20 \pm 0.30 & 0.058\pm 0.005 \\
            &  6 & 1.23\pm 0.32 & 0.052\pm 0.005
\enddata
\tablecomments{
The decorrelation bandwidth, $\Delta f_{1/e}$ and the magnitude of the maximum
frequency shift, with standard errors, along pulse longitude for scans 1 to 6
for each of the two observing dates. For the computation of the error of $\Delta
f_{1/e}$, see \citet{Bartel+2022}. For errors of the frequency shift, see
caption of Figure ~\ref{fig5}}
\label{tab:shift_vs_decorr}
\end{deluxetable*}
\section{Cause and Characterization of distortions}\label{sec:causes}
\subsection{Effect similar to dispersion delay} \label{subsec:delay}
What could be the cause of the observed phase rate and frequency shift changing
across the pulse profile?  Earlier, \citet{Bartel+1985} considered the varying
polarisation as a function of pulse longitude combined with polarisation
impurities of the feed that could lead to interferometer phase changes across
the pulse profile.  Here we will show that low-level digitization effects lead
to the distortions as a function of longitude that can be best illustrated by
considering the dispersion delay of the pulsar signal across the receiver
bandpass combined with the diffraction features of the dynamic spectrum before
dedispersion.

\begin{figure*}[htb]
\includegraphics{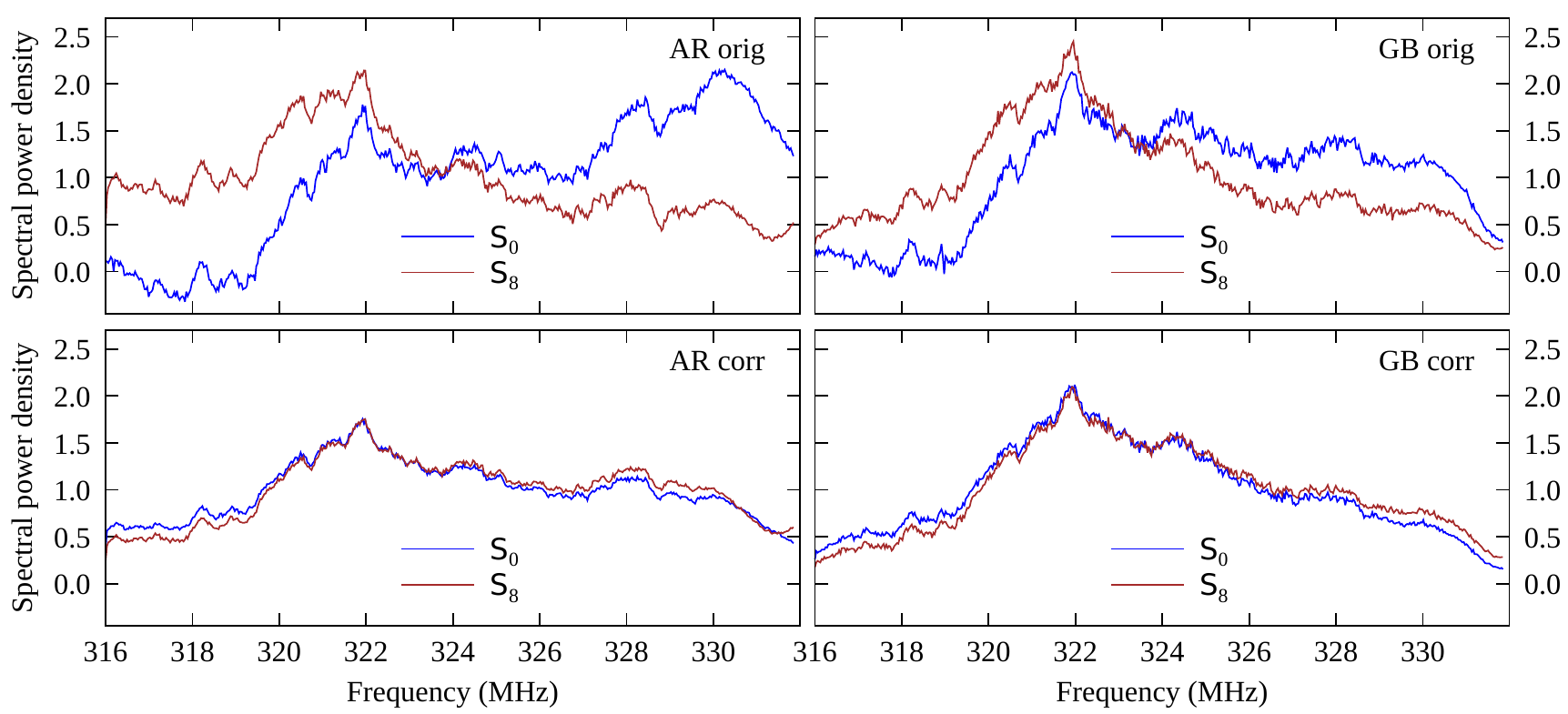}
\caption{Comparison of AR and GB time-averaged LCP spectra for leading ($S_0$)
  and trailing ($S_8$) windows, each with the spectrum for the off-pulse
  subtracted.  Upper panels show spectra before correction and lower panels
  after correction. The spectra correspond to scan 1 of 1160 s duration on 2017
  December 22. The spectra were normalized so that the frequency-averaged power
  spectral density $\bar S=1$.
\label{fig8} }
\end{figure*}

In Figure~\ref{fig8} (left upper panel) we present the dynamic spectra,
S$_0$ and S$_8$, for AR of Figure~\ref{fig3} (left panels) for the
leading and trailing windows respectively but averaged in time. For comparison
we also show the corresponding spectra for GB (right upper panel).  For each of
AR and GB the S$_0$ spectra of the leading window \win{0} appear to be shifted
toward higher frequencies compared to the S$_8$ spectra of the trailing window
\win{8}.

\begin{figure}[htb]
\includegraphics{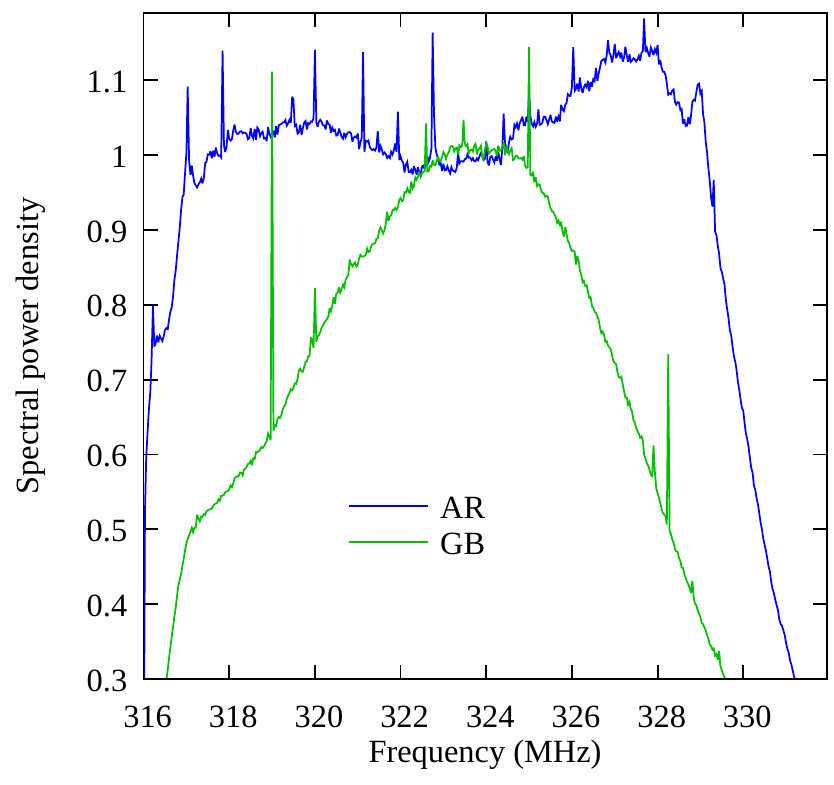}
\caption{Bandpasses for AR and GB from average spectra of off-pulse longitude
  windows, \win9 to $w_{11}$, from scan 1 on 2017 December 22.
\label{fig9}}
\end{figure}

These characteristics are similar to what could be expected for a pulsar signal
sweeping down in frequency with different pulse components illuminating scintles
in the spectrum at different frequencies.

The average profile of PSR B1237+25 consists of 5 components as it is shown in
Figure~\ref{fig1}.  With a dispersion measure, $DM=9.3~\text{cm}^{-3}~\text{pc}$, it takes
about 36~ms for the pulsar signal to sweep across the 16 MHz band from 332 to
316 MHz.  The spectrum $S_0$ corresponds to the case where the strong leading
component comes to the high frequency part of the bandpass dominating the
illumination and amplification of the diffraction features in the averaged
dynamic spectrum between 332 and 328 MHz, thus causing a false visible shift of
the spectrum to higher frequencies.  The weight of the illumination and the
amplification changes and becomes more balanced while the pulse travels through
the bandpass, minimizing the frequency shift in the central region of our
profile. However, the shift continues clearly further to lower frequencies when
the strong trailing component of the pulse profile dominates the illumination
and amplification of the dynamic spectrum between 324 and 318 MHz causing a
false visible shift of $S_8$ to lower frequencies.  This effect is already
indicated in Figure ~\ref{fig3} (left panels) for the dynamic spectra
of \win{0} and \win{8} but can be more clearly seen in Figure~\ref{fig8}.

Also, the effect is particularly strong for AR for which the bandpass is almost
flat across the 16 MHz bandwidth (see, Figure~\ref{fig9}). In contrast,
the bandpass for GB attenuates the high and low frequencies of the full
bandwidth which weakens the effect of the illuminations of the diffraction
features at the filter ends. The result is a smaller frequency shift which can
be seen in Figures~\ref{fig4}, \ref{fig6}, and
\ref{fig8}.

\subsection{Effect of signal digitization} \label{sub:digit}
The ASC correlator corrects for the dispersion delay by composing the full spectrum
for a given time from a sample of delayed spectra. In our case we used 1000 time
bins per pulsar period, giving us 1000 corresponding spectra with 512 channels
each. With this number of channels our correlator produces spectra every
$32~\mu$s. Each such spectrum is subject to a redistribution of harmonics between
corresponding bin spectra. Thus, for every pulsar period, the spectrum at every
bin will be corrected for the dispersion delay.  Our processing system is
described by \citet{2017JAI.....650004L}.  Nevertheless we still have, for
spectra at different longitudes, distortions similar to those expected due to
the influence of the dispersion delay. 

The reason for the distortions can be traced back to the effects of \nb{digitizing a non-stationary stochastic signal like the pulsar signals probed in this paper} \citep{Jenet1998}.  For one-bit digitizing
(clipping) where negative values are recorded as -1, and positive values as +1
the signal variance $\sigma^2=1.0$.  For a signal with frequency components, $1 \leq k\leq
N_f$, up to a maximum of $N_f$ the variance of a signal is related to the
spectral power density, $s_k$, by $\sigma^2=2\sum_{k=1}^{N_f} s_k$. If we assume that
there is an excess of spectral power density at relatively high frequencies
beyond component $a$, with $k>a$, then there will be a false deficiency of
spectral power density at relatively low frequencies, $k<a$. Let us assume that
the current output spectrum from the correlator has such high frequency excess
because the beginning of the pulse reached the receiver band.  This output
single spectrum would be redistributed between time bins with inadequate values,
namely, underestimated low frequency values of the spectral power density would
go to the preceding time bins causing a false decrease of total power.  On the
other hand, the high frequency portion of the spectrum at the leading longitude
of the average profile would be overestimated. Thus, the one-bit digitizing acts
like a non-compensated dispersion delay.

The same effect can be seen under saturation conditions for two-bit (four level)
digitizing used at both AR and GB.  Under normal conditions with no saturation
of the signal four values are utilized for such digitizing, -3, -1, +1, and
+3. A transition level, $s_0$, between $\pm 1$ and $\pm 3$ values must be close
to $\sigma$, with $s_0=0.995\sigma$ \citep{2017isra.book.....T}. An automatic gain control
system (AGC) is used at each of our telescopes to keep the system at such a
level during VLBI observing session. The AGC will compensate progressive slow
signal variations caused by a change in elevation of the source or weather
condition. For pulsar observations the AGC would not work correctly due to its
inertia. Therefore, we switched off the AGC system in our observations. With the
AGC switched off, the digitizer was saturated for strong pulses at such
sensitive radio telescopes as AR and GB.  Under saturation conditions the
digitizer acts like a one-bit sampler with only values of $\pm 3$, generating
false spectral distortions.  One can see this effect in the average profiles
shown in Figure~\ref{fig1} \nb{most clearly where the intensity dips below the baseline on each side of the profile.} Such digitizing also acts like a
non-compensated dispersion delay. For multilevel digitization of pulsar signals
the non-compensated dispersion delay is less likely and decreases with the
number of digitization levels.
\subsection{Distortion observations of other pulsars} \label{sub:1133}
\begin{figure}[htb]
\includegraphics{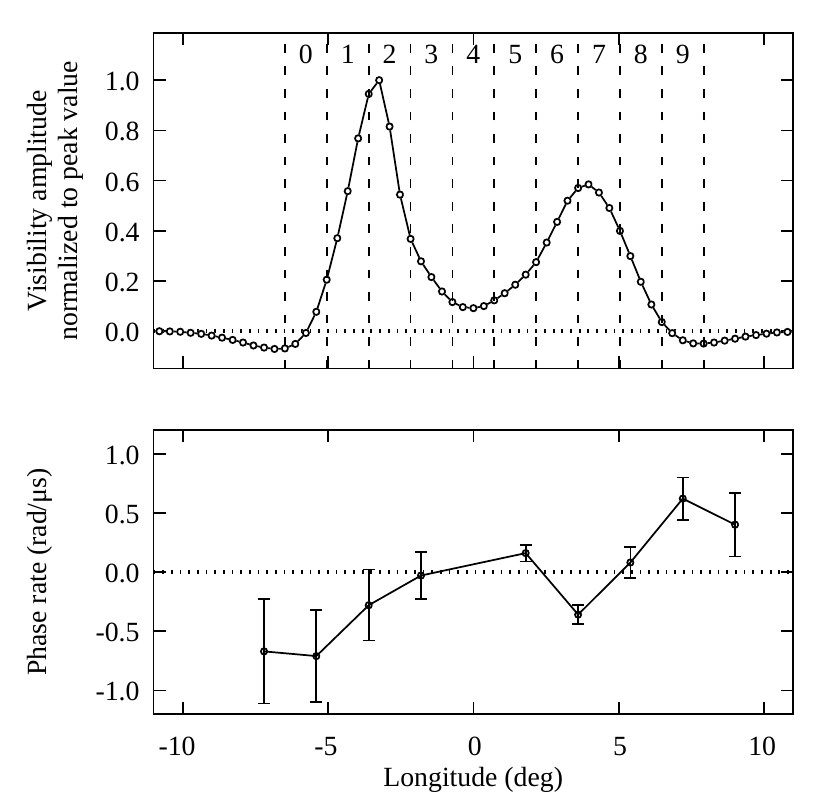}
\caption{Top: Average profile of PSR B1133+16 at LCP obtained at AR for \nb{scan 1}
  on 2018 December 17. Bottom: Phase rates as the
  derivatives of the AR-GB VLBI visibility phases with respect to delay as a
  function of pulse longitude for the same observing scan. \nb{Plots are based on original, uncorrected data.} For more information,
  compare with Figure~\ref{fig1}.
  \label{fig10}}
\end{figure}

Can the effect of distortion also be found in other pulsars?  As an example we
present in Figure~\ref{fig10} results for PSR B1133+16, with an average
profile with two components, $P=1.19$~s, $DM=4.84$~cm${}^{-3}$~pc, \nb{and $\Delta f_{1/e} = 1.8$~MHz.}  We analyzed the average profile and the phase rate
between visibility functions for different longitude windows, as described in
Section 3 and plot the results in Figure ~\ref{fig10}.
As for PSR B1237+25 the average pulse profile is distorted with intensity
decreases at both sides of the profile, and the phase rates vary
non-monotonically as a function of pulse longitude.  The maximum difference is
$\sim$1.2 rad/$\mu$s which corresponds to a frequency shift of $\sim$190 kHz. The shift
is again from higher frequencies at the leading part of the pulse profile to
lower frequencies at the trailing part. However, the magnitude of the shift for
this pulsar is much larger than what would be expected from Figure
~\ref{fig7} for PSR B1237+25. Clearly, the maximum frequency
shift across the pulse profile depends also on the pulsar under study.

In general we think that this distortion effect can be
observed in any pulsar independent of the complexity of its
profile. However while the frequency shift across a profile
with two or more components is rather modulated, this
frequency shift would likely be much smoother and generally
monotonic for a single component profile. 
\section{Significant reduction of distortion}
\label{sec:reduction}
As was indicated in the previous section, the observed distortions as a function
of pulsar longitude can be explained in terms of effects of dispersion and
low-level digitization. Since all VLBI observations of the RadioAstron mission
were archived, we accessed the original VLBI data and applied the corrections of
2-bit sampling as outlined by \citet{Jenet1998}. 

\begin{figure}[htb]
\includegraphics{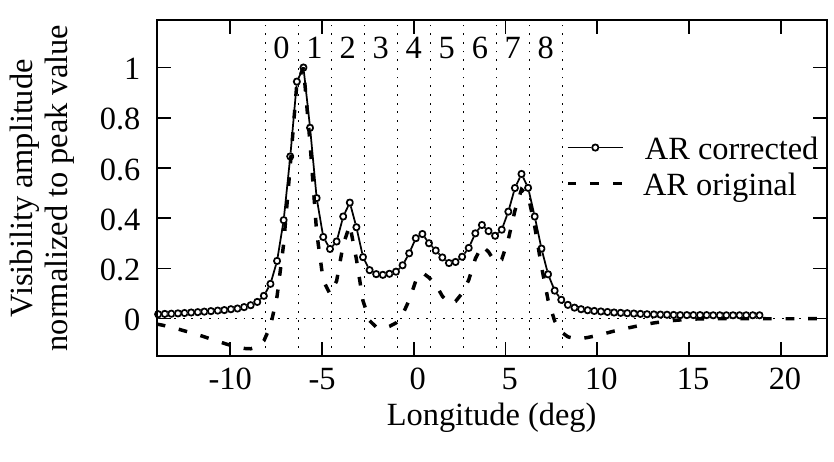}
\caption{Average profile of PSR B1237+25 \nb{for the original data obtained at AR for scan 1 on 2017 December 12 (copied from Figure 1) and the profile after correction. Note, the intensity dips on both sides of the profile have disappeared.} 
\label{fig11}}
\end{figure}

\nb{To give a short summary, the method of correction is based on using Gaussian statistics of samples of records of the original VLBI data before dedispersion. Without correction the average pulse profile displays characteristic intensity distortions including the dips before and after the profile (see Figure~\ref{fig11}) similar to the dips shown in \citep{Jenet1998} for the Vela pulsar under study by them. For the correction we first analyzed a very small part of the recorded signal and calculated the portion of samples $\psi$ corresponding to +/- 1 values. If there were no digitization distortions, this portion would be equal to 2/3 and corresponded to a correct threshold level $H$ of $\pm 1\sigma$ of the recorded signal. On strong pulses the statistics would be different, and with the calculated current portion $\psi$ the true value of the threshold, $\pm H$ of the distorted digitized data could be estimated for Gaussian distribution. The next step is to calculate the corrected values of the samples as a mean for Gaussian distribution between zero and $\pm H$ instead of $\pm 1$ and as a mean between $\pm H$ and $\pm\infty$ instead of $\pm 3$.  Thus, we corrected every sample using moving average for bit statistics. For another description of the application of digitization correction, see \citet{vStraten+2011}}.


We also coherently dedispersed the data \citep[see,][]{Hank1974} for improvement
over incoherent dedispersion although without any apparent effect on the
correction itself.  For \nb{further} details of our data reduction and
correction process, see \citet{Girin+2023}. In Figure ~\ref{fig11} we also show
the corrected dedispersed average profile for easy comparison with the
uncorrected profile. It can be clearly seen that the intensity distortions
including the dips outside the profile have disappeared.

%
The dips of intensity below the
off-pulse level on each side of the profile have disappeared.  As for the dynamic
spectra in \win0 and \win8, we show in Figure~\ref{fig3} that after
correction the frequency shift has also largely disappeared. The details of the
frequency shift as a function of longitude after correction are displayed in
Figure~\ref{fig5}.  The corrected frequency shifts (solid lines)
for both AR and GB are significantly reduced in amplitude and are almost
constant along longitude in comparison to the original data (dashed lines). The
same characteristics can also be found in the averaged spectra in
Figure~\ref{fig6}.  

Further insight can be derived from the
time-averaged spectra in Figure~\ref{fig8}. Close inspection shows that the
narrow spectral features in the original data are not shifted along
longitude. Instead the weight within the spectra is shifted. For the leading
part of the pulse the spectrum, $S_0$, has on average more power at higher
frequencies which shifts to lower frequencies in the spectrum for the trailing
part of the profile, $S_8$ (see also interpretation in
subsection~\ref{subsec:delay}).  Again, the corrected spectra show almost no
differences between those of the leading and trailing pulse parts.
\section{Discussion}
\label{sec:discussion}
Our observations can be compared with those of PSRs B1237+25 at 430 MHz at AR by
\citet{wol_cor1987} and B1133+16 at 327 MHz at Ooty by \citet{GuptaBR1999}. \nb{These
authors observed that the dynamic spectra as a function of longitude were shifted non-monotonically from high frequencies at the leading part of the pulse profile to low
frequencies at the trailing part within their receiver bandpasses.}

We found a similar shift of the spectra from high to low frequencies in our VLBI
observations with AR and GB as well as in observations with each single
telescope.  In particular, \citet{wol_cor1987} reported for PSR B1237+25 a
frequency shift of 39 kHz across the pulse profile. The decorrelation bandwidth
for their two days of observations was 442 and 615 kHz. Although their observing
frequency of 430 MHz was somewhat higher than ours, their observed
maximum frequency shift is comparable with our prediction of ~10 kHz \nb{for 327 MHz} from Figure
\ref{fig7} which further indicates that the nature of their
longitudinal frequency shift of the dynamic spectra is similar to ours.

An important aspect is that both groups used, as we did, low-level digitizers,
at AR 3-level \citep{wol_cor1987} and at Ooty 1-bit samplers
\citep{Bhat+1999}. Through our analysis we can now trace our longitudinal
frequency shift back to digitization effects. In view of interstellar
interferometry we expect that most if not all pulsar observations with low-level
digitization, \nb{technically or because of saturation,} can be affected by longitudinal distortion if the effects discussed
above are not addressed in detail.

Despite having largely corrected for the low-level digitizing effects for PSR B1237+25, some
apparently significant residuals remain in our data. \nb{Are these due to small technical effects that still remain uncorrected or do they have an astrophysical origin and indicate a marginal resolution of the magnetosphere?} 

The cause is not
clear. Perhaps our estimated uncertainties are too small and the differences from zero frequency shifts
are not significant. The correction process itself may not be completely
effective for our data. Lastly, the polarization impurities of the telescope
feeds, although not subject of this paper, need to be considered when computing
the limiting effect of the changing polarization characteristics along pulsar
longitude on the phase and frequency shifts \citep[see,][]{Bartel+1985}. In this
respect, the frequency shifts along longitude in RCP and LCP for AR and GB (see,
Figure~\ref{fig1}) are of interest. While there is fair consistency
between the curves, there are also slight differences between the RCP and LCP
curves that vary with longitude differently for AR and GB.  Such an effect may
be conceivable for different polarization impurities of the feeds at the two
telescopes.

\nb{If however, the discrepancies have an astrophysical cause, then it is possible to elaborate on the emission separations and corresponding altitudes for a pulsar magnetic dipole field.
In this context it is interesting to note that, as displayed in 
Figure \ref{fig6}, after correction the differences in the frequency shifts between the leading and trailing parts of the profiles where the shifts are largest, averaged for AR and GB, are about 100 kHz on Dec. 2017 but only 10 kHz on February 2018. In relative terms the remaining shifts are about 15 and 25$\%$ of the uncorrected shifts, respectively. If the remaining frequency shifts of the corrected data had an astrophysical origin, then the shifts would be expected to be of the same magnitude on both days. That they were not and instead have similar portions of the shifts of their uncorrected data suggests that a large part of the residual shifts are still due to remaining uncorrected technical effects as discussed above. 
Any residual frequency shift variations due to having resolved the magnetosphere must therefore be smaller than even a portion of the largest shift variations of the corrected data from February 2018 in Figure \ref{fig6}.}

\nb{\citet{wol_cor1987} estimated for PSR B1237+25 emission separations of $\sim$1,000 km and respective altitudes of $\rLC$. If we take for our data an upper limit of any astrophysics related freqency shift of, say $10\%$ of the uncorrected frequency shifts, and apply this correction factor to the data of \citet{wol_cor1987}, then their estimate of the emission altitude would be only $0.1\rLC$, eliminating any need for a very distorted dipole field or a screen very close to the pulsar.
However, as for deriving estimates directly from our observations, their computations for refractive scintillation would not be directly applicable to our data with diffractive scintillation.} 

\nb{Diffractive scintillation was studied by \citet{cwb1983} and applied to a correlation analysis of dynamic spectra of PSRs B0525+21 and B1133+16 with AR at 430 MHz. No correlation degradation or significant frequency shifts beyond the spectral resolution of 1.2 and 40 kHz for the two pulsars respectively, were found for the spectra of the leading and trailing parts of the pulse profiles. With their model they derived emission separations of also $\sim1,000$~km and emission altitudes of $<0.06\rLC$ and $<0.53\rLC$. From these comparisons our upper limit of only a couple of kHz of astrophysics related frequency shift would suggest that the emission altitude is also well below $\rLC$.}

In general, we think that the constraints discussed here need to be taken into account for
observations with the goal of resolving spatially the pulsar magnetosphere. With
appropriate considerations a distortion-less use of interstellar interferometry
could likely be achieved.
\section{Summary and Conclusions}
\label{sec:conclusion}
Here we summarize our observations and give our conclusions.
\begin{enumerate}
\item
Inspired by earlier reports of having resolved the magnetosphere of pulsars with
interstellar interferometry, we used VLBI observations of PSR B1237+25 conducted
with AR and GB at 324 MHz and analyzed the interferometry as well as the single
telescope data.  All observations were done in the context of the RadioAstron
space-VLBI mission.
\item  
During a time of diffractive scintillation, the dynamic spectra changed as a
function of pulsar longitude with the spectrum of the leading part at higher
frequencies shifting for the trailing part to lower frequencies.
\item
In VLBI data as well as single telescope data the frequency shift displayed a
\nb{non-monotonic} pattern as a function of pulsar longitude. Although the
patterns were largely similar for AR and GB, differences could be due to
differences in the bandpasses.
\item
The maximum frequency shift, between the spectra of the leading and trailing
parts of the pulse profile is a steep function of decorrelation bandwidth.
\item
The integrated pulse profile showed characteristic deficiencies in intensity at
both sides of the profile.
\item
Similar distortions of the scintillation spectra as a function of longitude and
of the pulse profile were found for PSR B1133+16.
\item
Despite having observed characteristic phase and frequency shifts in
scintillation spectra as a function of pulsar longitude, we do not contribute
the shifts to having resolved the pulsar magnetosphere.
\item
We attribute the distortions and frequency shifts of the longitude-dependent
dynamic spectra mostly to uncorrected low-level digitizing of the data.
\item
With the data corrected \citep[see,][]{Jenet1998}, the \nb{non-monotonic}
frequency shift pattern of the dynamic spectra along pulse longitude largely
disappeared.
\item
Small remaining distortions could perhaps partly be caused by polarization
impurities of the feeds.
\item 
\nb{Upper limits of any astrophysics related frequency shifts would suggest that the altitude of emission regions for PSR B1237+25 is well below $\rLC$.}
\item
In view of our analysis we think that observations with the intend to resolve
the pulsar magnetosphere need to be critically evaluated in terms of these
constraints on interstellar interferometry.
\end{enumerate}
\begin{acknowledgments}
The RadioAstron project is led by the Astro Space Center of the
Lebedev Physical Institute of the Russian Academy of Sciences and the
Lavochkin Scientific and Production Association under a contract with the
Russian Federal Space Agency, in collaboration with partner organizations in
Russia and other countries. N.B. was supported by the national Sciences and
Engineering Research Council of Canada.  
The Arecibo Observatory is a facility
of the National Science Foundation operated under cooperative agreement by the
University of Central Florida and in alliance with Universidad Ana G. Mendez,
and Yang Enterprises, Inc.  The Green Bank Observatory is a facility of the
National Science Foundation operated under cooperative agreement by Associated
Universities, Inc.
\end{acknowledgments}
\facilities{Arecibo, Green Bank Telescope}.
\bibliographystyle{aasjournal}
\bibliography{MVPopov}

\begin{thebibliography}{}
\expandafter\ifx\csname natexlab\endcsname\relax\def\natexlab#1{#1}\fi
\providecommand{\url}[1]{\href{#1}{#1}}
\providecommand{\dodoi}[1]{doi:~\href{http://doi.org/#1}{\nolinkurl{#1}}}
\providecommand{\doeprint}[1]{\href{http://ascl.net/#1}{\nolinkurl{http://ascl.net/#1}}}
\providecommand{\doarXiv}[1]{\href{https://arxiv.org/abs/#1}{\nolinkurl{https://arxiv.org/abs/#1}}}

\bibitem[{{Backer}(1975)}]{Backer1975}
{Backer}, D.~C. 1975, \aap, 43, 395

\bibitem[{{Bartel} {et~al.}(2022){Bartel}, {Burgin}, {Fadeev}, {Popov},
  {Ronaghikhameneh}, {Smirnova}, \& {Soglasnov}}]{Bartel+2022}
{Bartel}, N., {Burgin}, M.~S., {Fadeev}, E.~N., {et~al.} 2022, \apj, 941, 112,
  \dodoi{10.3847/1538-4357/ac9eae}

\bibitem[{{Bartel} {et~al.}(1985){Bartel}, {Ratner}, {Shapiro}, {Cappallo},
  {Rodgers}, \& {Whitney}}]{Bartel+1985}
{Bartel}, N., {Ratner}, M.~I., {Shapiro}, I.~I., {et~al.} 1985, \aj, 90, 2532,
  \dodoi{10.1086/113958}

\bibitem[{{Bhat} {et~al.}(1999){Bhat}, {Rao}, \& {Gupta}}]{Bhat+1999}
{Bhat}, N.~D.~R., {Rao}, A.~P., \& {Gupta}, Y. 1999, \apjs, 121, 483,
  \dodoi{10.1086/313198}

\bibitem[{{Brisken} {et~al.}(2010){Brisken}, {Macquart}, {Gao}, {Rickett},
  {Coles}, {Deller}, {Tingay}, \& {West}}]{Brisken+2010}
{Brisken}, W.~F., {Macquart}, J.~P., {Gao}, J.~J., {et~al.} 2010, \apj, 708,
  232, \dodoi{10.1088/0004-637X/708/1/232}

\bibitem[{{Cordes} {et~al.}(1986){Cordes}, {Pidwerbetsky}, \&
  {Lovelace}}]{CordesPidweb1986}
{Cordes}, J.~M., {Pidwerbetsky}, A., \& {Lovelace}, R.~V.~E. 1986, \apj, 310,
  737, \dodoi{10.1086/164728}

\bibitem[{{Cordes} {et~al.}(1983){Cordes}, {Weisberg}, \&
  {Boriakoff}}]{cwb1983}
{Cordes}, J.~M., {Weisberg}, J.~M., \& {Boriakoff}, V. 1983, \apj, 268, 370,
  \dodoi{10.1086/160961}

\bibitem[{{Girin} {et~al.}(2023){Girin}, {Likhachev}, {Rudnitskiy},
  {Andrianov}, [Burgin], {Popov}, {Soglasnov}, \& {Zuga}}]{Girin+2023}
{Girin}, I.~A., {Likhachev}, S.~F., {Rudnitskiy}, A.~G., {et~al.} 2023,
  {Processing system for coherent dedispersion of pulsar emission}.
\newblock \doarXiv{2303.17280}

\bibitem[{{Gupta} {et~al.}(1999){Gupta}, {Bhat}, \& {Rao}}]{GuptaBR1999}
{Gupta}, Y., {Bhat}, N.~D.~R., \& {Rao}, A.~P. 1999, \apj, 520, 173,
  \dodoi{10.1086/307442}

\bibitem[{{Gwinn} {et~al.}(1998){Gwinn}, {Britton}, {Reynolds}, {Jauncey},
  {King}, {McCulloch}, {Lovell}, \& {Preston}}]{1998ApJ...505..928G}
{Gwinn}, C.~R., {Britton}, M.~C., {Reynolds}, J.~E., {et~al.} 1998, \apj, 505,
  928, \dodoi{10.1086/306178}

\bibitem[{{Hankins}(1974)}]{Hank1974}
{Hankins}, T.~H. 1974, \aaps, 15, 363

\bibitem[{{Jenet} \& {Anderson}(1998)}]{Jenet1998}
{Jenet}, F.~A., \& {Anderson}, S.~B. 1998, \pasp, 110, 1467,
  \dodoi{10.1086/316273}

\bibitem[{{Johnson} {et~al.}(2012){Johnson}, {Gwinn}, \&
  {Demorest}}]{2012ApJ...758....8J}
{Johnson}, M.~D., {Gwinn}, C.~R., \& {Demorest}, P. 2012, \apj, 758, 8,
  \dodoi{10.1088/0004-637X/758/1/8}

\bibitem[{{Likhachev} {et~al.}(2017){Likhachev}, {Kostenko}, {Girin},
  {Andrianov}, {Rudnitskiy}, \& {Zharov}}]{2017JAI.....650004L}
{Likhachev}, S.~F., {Kostenko}, V.~I., {Girin}, I.~A., {et~al.} 2017, Journal
  of Astronomical Instrumentation, 6, 1750004,
  \dodoi{10.1142/S2251171717500040}

\bibitem[{{Lovelace}(1970)}]{1970PhDT.......113L}
{Lovelace}, R.~V.~E. 1970, PhD thesis, -

\bibitem[{{Main} {et~al.}(2021){Main}, {Lin}, {van Kerkwijk}, {Pen},
  {Rudnitskii}, {Popov}, {Soglasnov}, \& {Lyutikov}}]{Main2021}
{Main}, R., {Lin}, R., {van Kerkwijk}, M.~H., {et~al.} 2021, \apj, 915, 65,
  \dodoi{10.3847/1538-4357/ac01c6}

\bibitem[{{Pen} {et~al.}(2014){Pen}, {Macquart}, {Deller}, \& {Brisken}}]{Pen}
{Pen}, U.-L., {Macquart}, J.-P., {Deller}, A.~T., \& {Brisken}, W. 2014,
  \mnras, 440, L36, \dodoi{10.1093/mnrasl/slu010}

\bibitem[{{Rickett}(1977)}]{1977ARA&A..15..479R}
{Rickett}, B.~J. 1977, \araa, 15, 479,
  \dodoi{10.1146/annurev.aa.15.090177.002403}

\bibitem[{{Scheuer}(1968)}]{1968Natur.218..920S}
{Scheuer}, P.~A.~G. 1968, \nat, 218, 920, \dodoi{10.1038/218920a0}

\bibitem[{{Smirnova}(1992)}]{smirnova1992}
{Smirnova}, T.~V. 1992, Soviet Astronomy Letters, 18, 392

\bibitem[{{Smirnova} \& {Shishov}(1989)}]{SS1989}
{Smirnova}, T.~V., \& {Shishov}, V.~I. 1989, Soviet Astronomy Letters, 15, 191

\bibitem[{{Smirnova} {et~al.}(1996){Smirnova}, {Shishov}, \&
  {Malofeev}}]{SSM1996}
{Smirnova}, T.~V., {Shishov}, V.~I., \& {Malofeev}, V.~M. 1996, \apj, 462, 289,
  \dodoi{10.1086/177150}

\bibitem[{{Thompson} {et~al.}(2017){Thompson}, {Moran}, \&
  {Swenson}}]{2017isra.book.....T}
{Thompson}, A.~R., {Moran}, J.~M., \& {Swenson}, George~W., J. 2017,
  {Interferometry and Synthesis in Radio Astronomy, 3rd Edition} (Springer,
  Cham), \dodoi{10.1007/978-3-319-44431-4}

\bibitem[{{van Straten} \& {Bailes}(2011)}]{vStraten+2011}
{van Straten}, W., \& {Bailes}, M. 2011, \pasa, 28, 1, \dodoi{10.1071/AS10021}

\bibitem[{{Wolszczan} \& {Cordes}(1987)}]{wol_cor1987}
{Wolszczan}, A., \& {Cordes}, J.~M. 1987, \apjl, 320, L35,
  \dodoi{10.1086/184972}

\end{thebibliography}
\end{document}